\documentclass[10pt,journal,twocolumn]{IEEEtran}
\IEEEoverridecommandlockouts
\usepackage{cite}
\usepackage{amsmath,amssymb,amsfonts,booktabs}
\usepackage{graphicx}
\usepackage{textcomp}
\usepackage{xcolor}
\usepackage{enumerate}
\usepackage{epstopdf}
\usepackage{subcaption}
\usepackage{amsthm}

\usepackage{algorithm}
\usepackage{algorithmic}

\allowdisplaybreaks[4]

\begin{document}

\title{\LARGE UAV-Enabled Wireless Networks with Movable-Antenna Array:\\Flexible Beamforming and Trajectory Design}

\vspace{-0.5em}
\author{Wenchao Liu,~\IEEEmembership{Student Member,~IEEE},
        Xuhui Zhang,~\IEEEmembership{Student Member,~IEEE},
        Huijun Xing,~\IEEEmembership{Student Member,~IEEE},\\
        Jinke Ren,~\IEEEmembership{Member,~IEEE},
        Yanyan Shen,~\IEEEmembership{Member,~IEEE},
        and Shuguang Cui,~\IEEEmembership{Fellow,~IEEE}

% \thanks{This work was supported in part by the Shenzhen Science and Technology Program under Grant JCYJ20220818101607015, the Foundation of Key Laboratory of System Control and Information Processing, Ministry of Education, Shanghai, China under Grant Scip20240114, the National Natural Science Foundation of China under Grant 61503368,
% in part by the Basic Research Project of Hetao Shenzhen-HK S\&T Cooperation Zone under Grant HZQB-KCZYZ-2021067, the NSFC under Grant 62293482, the Shenzhen Outstanding Talents Training Fund under Grant 202002, the Guangdong Research Projects under Grant 2017ZT07X152 and 2019CX01X104, the Guangdong Provincial Key Laboratory of Future Networks of Intelligence under Grant 2022B1212010001, and the Shenzhen Key Laboratory of Big Data and Artificial Intelligence under Grant ZDSYS201707251409055.
% (\emph{Wenchao Liu and Xuhui Zhang contributed equally to this work.})
% (\emph{Corresponding author: Yanyan Shen.})
% }

\thanks{Copyright (c) 2024 IEEE.
Personal use of this material is permitted.
However, permission to use this material for any other purposes must be obtained from the IEEE by sending a request to pubs-permissions@ieee.org.
}

\thanks{
W. Liu is with the Shenzhen Institute of Advanced Technology (SIAT), Chinese Academy of Sciences, Guangdong 518055, China, and is also with the Southern University of Science and Technology, Guangdong 518055, China (e-mail: wc.liu1@siat.ac.cn).
}

\thanks{
X. Zhang and J. Ren are with the Shenzhen Future Network of Intelligence Institute (FNii-Shenzhen), the School of Science and Engineering (SSE), and the Guangdong Provincial Key Laboratory of Future Networks of Intelligence, The Chinese University of Hong Kong, Shenzhen, Guangdong 518172, China (e-mail: xu.hui.zhang@foxmail.com; jinkeren@cuhk.edu.cn).
}

\thanks{
H. Xing is with the Department of Electrical and Electronic Engineering, Imperial College London, London SW7 2AZ, The United Kingdom, and is also with the FNii-Shenzhen and the Guangdong Provincial Key Laboratory of Future Networks of Intelligence, The Chinese University of Hong Kong, Shenzhen, Guangdong 518172, China (e-mail: huijunxing@link.cuhk.edu.cn).
}

\thanks{
Y. Shen is with the SIAT, Chinese Academy of Sciences, Guangdong 518055, China (e-mail: yy.shen@siat.ac.cn).
}

\thanks{
S. Cui is with the SSE, the FNii-Shenzhen, and the Guangdong Provincial Key Laboratory of Future Networks of Intelligence, The Chinese University of Hong Kong, Shenzhen, Guangdong 518172, China (e-mail: shuguangcui@cuhk.edu.cn).
}

% \thanks{
% Notice: This work has been submitted to the IEEE for possible publication. Copyright may be transferred without notice, after which this version may no longer be accessible.
% }
\vspace{-2.6em}
}

\maketitle

\begin{abstract}
Recently, movable-antenna (MA) array becomes a promising technology for improving the communication quality in wireless communication systems.
In this letter, an unmanned aerial vehicle (UAV) enabled multi-user multi-input-single-output system enhanced by the MA array is investigated. To enhance the throughput capacity, we aim to maximize the achievable data rate by jointly optimizing the transmit beamforming, the UAV trajectory, and the positions of the MA array antennas.
The formulated data rate maximization problem is a highly coupled non-convex problem,
for which an alternating optimization based algorithm is proposed to get a sub-optimal solution.
Numerical results have demonstrated the performance gain of the proposed method compared with conventional method with fixed-position antenna array.
\end{abstract}

\section{Introduction}
\IEEEPARstart{R}{ecently}, unmanned aerial vehicle (UAV)-enabled communications have emerged as a key technology to ensure long-term communications and flexible placements in the next generation cellular networks
\cite{8918497}.
The working altitude and mobility of the UAVs can provide stronger line-of-sight links among the UAVs and the ground users compared with conventional ground base stations, which improves the channel 
quality for the intensive demand of users \cite{9456851}.
However, the achievable data rate is still a bottleneck in UAV-enabled communications.
{To address this issue, movable-antenna (MA) technology is proposed to provide flexible beamforming to improve the communication throughout} \cite{10286328}.
{Specifically, the flexible beamforming is achieved by adjusting the positions of the antennas at the transmitter or/and the receiver in their certain regions to enable better quality of communication links.} Thus, the higher throughput capacity can be obtained \cite{10480333}.

The effectiveness, robustness, and adaptability of the MA array have been verified in previous works \cite{10236898, 10278220, 10416363, 10243545, 10354003}.
Specifically, compressed sensing based channel estimation for wireless communications enhanced with MA array is investigated in \cite{10236898}.
The simultaneously enhancing and/or nulling the signal power at desired and/or undesired directions with MA array is studied in \cite{10278220}.
In \cite{10416363}, the achievable secrecy rate is optimized in the MA array-enabled secure communication system with joint optimization of transmit beamforming and the positions of all movable antennas.
The transmit covariance matrix and the position of each MAs are jointly designed for a multiple-input multiple-output system in \cite{10243545}.
Moreover, the total transmit power of users is optimized in an MA-enhanced multiple-access channel system studied in \cite{10354003}.
On the other hand, the UAV-assisted communications have shown to improve the communication efficiency \cite{8663615, 9273074, 10054450, 10214196}.
The time allocation and UAV trajectory are joint optimized
to minimize the system energy consumption \cite{8663615}.
The minimum data rate of Internet of thing (IoT) users is maximized in
an IoT system studied in \cite{9273074}, where multiple UAVs are dispatched to power the IoT users, and then to collect data transmitted by the IoT users. 
To better represent the data rate and timeliness, a task-aware UAV trajectory is designed in \cite{10054450}, where the age of information is minimized among all ground users.
In \cite{10214196}, the non-orthogonal multiple access enabled UAV systems is investigated, where the energy efficiency is maximized among all users to improve spectrum efficiency and large-scale connectivity.
However, in previous works \cite{10236898, 10278220, 10416363, 10243545, 10354003}, the MA arrays are only deployed at fixed entity, lacking of flexibility.
Moreover, these UAV-enabled systems \cite{8663615, 9273074, 10054450, 10214196} only consider a single antenna or a fixed antenna array. 
{Therefore, they are short of
the utilization of flexible beamforming towards users, resulting in throughput bottleneck}.

{To solve these problems, we aim to fully utilize the flexible beamforming benefited by the design of the MA array according to the positions of the UAV and users in a multi-user (MU) multi-input-single-output (MISO) system.}
The total achievable data rate of all users is maximized
by jointly optimizing the transmit beamforming, the positions of all MAs, and the flying trajectory of the UAV. An efficient alternating optimization based algorithm is developed to solve the non-convex problem. Simulation results demonstrate the performance gain of proposed scheme compared to the conventional fixed-position antenna array scheme.

\section{System model and Problem Formulation}
We consider a UAV communication system, where a single UAV serves $K$ single-antenna users. The set of users is denoted as $\mathcal{K} \triangleq \{1, \cdots, K\}$.
{The UAV flies from the initial position to the final position within a time period of $T$ seconds. The time period is divided into $N$ slots, each with duration $\tau = \frac{T}{N}$.
The UAV provides data transmission services for all users at each time slot.}
Since the flying routine of the UAV is continuous, we discretize the flying by a certain position in each time slot, where the
two-dimensional (2D) location of the UAV at time slot $n$ is given by $\boldsymbol{q}[n] = [q_x[n], q_y[n]]$, with fixed altitude $H$.
Hence, the velocity of the UAV is $\boldsymbol{v}[n] = \frac{1}{\tau}(\boldsymbol{q}[n] - \boldsymbol{q}[n-1])$, which is limited by the minimum speed $V_{\min}$ and the maximum speed $V_{\max}$.
Besides, the acceleration of the UAV $\boldsymbol{a}[n] = \frac{1}{\tau}(\boldsymbol{v}[n] - \boldsymbol{v}[n-1])$ is also bounded by the maximum acceleration $a_{\max}$.
The location of the $k$-th user is denoted as $\boldsymbol{s}_k = [s_{k,x}, s_{k,y}]$.

{The UAV is equipped with a linear MA array with size $M$, where the number of antennas is not smaller than the number of users, i.e., $M \geq K$}.
The set of the MA array is denoted as $\mathcal{M} \triangleq \{1, \cdots, M\}$, where the position of the $m$-th MA at time slot $n$ is given by $x_m [n]$.
{Hence, the positions of the MA array can be written as $\boldsymbol{x}[n] = [x_1[n],\cdots,x_M[n]]^{\mathrm{T}}$.
To facilitate the following analysis, it is assumed that $0 \le x_{1}[n] \le x_{2}[n] \le \cdots \le x_{M}[n] \le L$, where $L$ is the maximum movable range of antennas}. Thus, the steering vector of the MA array can be expressed as
\begin{equation}{ \footnotesize
\begin{split}
    \boldsymbol{a}(\boldsymbol{x}[n],\theta_k [n]) = \left [
    e^{\mathrm{j}\frac{2\pi}{\lambda}x_1[n]\cos\theta_k [n]},
    \cdots,
    e^{\mathrm{j}\frac{2\pi}{\lambda}x_M[n]\cos\theta_k [n]}
    \right ]^{\mathrm{T}}, \label{steering vector}
\end{split} }
\end{equation}
where $\theta_k [n] = \arccos \frac{H}{\sqrt{\Vert \boldsymbol{q}[n]-\boldsymbol{s}_k\Vert^2+H^2}}$ is the steering angle towards the $k$-th user and $\lambda$ is the wavelength. We use $\boldsymbol{a}_k[n]$ to represent $\boldsymbol{a}(\boldsymbol{x}[n],\theta_k [n])$ for simplicity.
Let $\boldsymbol{w}_k [n] \in \mathbb{C}^{M\times1}$ denote the transmit beamforming vector of the UAV for communicating with the $k$-th user at time slot $n$.
We assume the UAV-to-ground links are line of sight (LoS) channels\footnote{
{
Owing to the flexibility of the UAV, it can fly over obstacles and get close to users to achieve the LoS channel conditions \cite{9916163}.
}}, and the free-space path loss from the UAV to the $k$-th user is given by $h_k[n] = \sqrt{\chi} \left(\Vert \boldsymbol{q}[n]-\boldsymbol{s}_k\Vert^2+H^2 \right)^{-\frac{1}{2}}$, where $\chi$ is a constant value at a reference distance $1 \rm{m}$.
To this end, the received signal-to-interference-plus-noise ratio (SINR) of the $k$-th user at time slot $n$ can be given by
\begin{equation} {\footnotesize
\begin{split}
        \gamma_k[n] (\boldsymbol{a}_k, \boldsymbol{w}_k, \boldsymbol{q}) = \frac{ \left \vert h_k[n]\boldsymbol{a}_k^{\mathrm{H}}[n]
     \boldsymbol{w}_k [n] \right \vert^{2} }
    {\sum_{l\in\mathcal{K}\backslash\{k\} } \vert 
    h_k[n]\boldsymbol{a}_k^{\mathrm{H}}[n] \boldsymbol{w}_l[n] \vert^{2} + \sigma_k^2},
\end{split} }
\end{equation}
where $\sigma_k^2$ denotes the noise power at the $k$-th user. The achievable data rate
% in bps/Hz
of the UAV for communicating with the $k$-th user at time slot $n$ can be given by
\begin{equation} {\small
    r_k[n](\boldsymbol{a}_k, \boldsymbol{w}_k, \boldsymbol{q}) = \log_2 (1+\gamma_k[n](\boldsymbol{a}_k, \boldsymbol{w}_k, \boldsymbol{q})). \label{rate}}
\end{equation}

In this work, we aim to maximize the total achievable data rate of users, by jointly optimizing the positions of the MA array $\boldsymbol{X} \triangleq \{\boldsymbol{x}[n], \forall n\}$, the transmit beamforming $\boldsymbol{W} \triangleq \{\boldsymbol{w}_k [n], \forall k, \forall n \}$ at the UAV, and the UAV trajectory $\boldsymbol{Q} \triangleq \{\boldsymbol{q}[n], \forall n\}$. Therefore, the problem is formulated as
\begin{subequations} { \small
\begin{flalign}
 (\textbf{P1}):\ \max_{\boldsymbol{X},\boldsymbol{W},\boldsymbol{Q}} \quad & \sum_{n= 1}^{N}\sum_{k= 1}^{K}  r_k[n](\boldsymbol{a}_k, \boldsymbol{w}_k, \boldsymbol{q})  \nonumber\\
 {\rm{s.t.}}  \quad & \sum_{k= 1}^{K} \Vert \boldsymbol{w}_k[n] \Vert^2 \le P_{\max},\ \forall n\in \mathcal{N},   \label{p1a}\\
 & \{x_m[n]\}_{m=1}^M \in [0,L],\ \forall n\in \mathcal{N},\label{p1b}\\
 &{ x_{p} [n] - x_{q} [n]  \geq d_{\min}, \ \text{where }\{p, q\} \in \mathcal{M}} , \nonumber\\
 &\quad\quad\quad\quad\quad\quad\quad\quad\quad\quad\quad
 { p > q,\ \forall n\in \mathcal{N} },\label{p1c}\\
 &V_{\min} \leq \Vert \boldsymbol{v}[n] \Vert \leq V_{\max},\ \forall n\in \mathcal{N}\backslash\{1\},\label{p1d}\\
 &0\leq \Vert \boldsymbol{a}[n] \Vert \leq a_{\max},\ \forall n\in \mathcal{N}\backslash\{1,2\},\label{p1e}
\end{flalign}}
\end{subequations}

\noindent where (\ref{p1a}) denotes the transmit power limitation, (\ref{p1b}) is the range of positions of all antennas, (\ref{p1c}) represents that the distance between adjacent MAs cannot lower than the minimum distance $d_{\min}$ for preventing the coupling effect, (\ref{p1d}) and (\ref{p1e}) are the UAV kinematic constraints. (\textbf{P1}) is non-convex due to the high coupling among the optimization variables, which is hard to solve by conventional methods. Conversely, alternating optimization (AO) is an efficient and robustness method to solve such problems, by iteratively solving part of the variables while fixing others. The AO-based algorithm for solving (\textbf{P1}) is presented in the following section.

\section{Alternating Optimization for (\textbf{P1})}
\subsection{Optimizing $\boldsymbol{W}$ Given $\boldsymbol{X}$ and $\boldsymbol{Q}$}
Let $\boldsymbol{W}_k[n] = \boldsymbol{w}_k[n]\boldsymbol{w}_k^{\mathrm{H}}[n]$, then we have $\boldsymbol{W}_k[n] \succeq \boldsymbol{0}$, and $\mathrm{rank}(\boldsymbol{W}_k[n] ) \leq 1$. By substituting $\boldsymbol{w}_k[n]\boldsymbol{w}_k^{\mathrm{H}}[n]$ with $\boldsymbol{W}_k[n]$, problem (\textbf{P1}) can be rewritten as
\begin{subequations} {\small
\begin{flalign}
 (\textbf{P2}):\ \max_{\{\boldsymbol{W}_k[n] \succeq 0\}} \quad & \sum_{n= 1}^{N}\sum_{k= 1}^{K}  \Tilde{r}_k[n] (\boldsymbol{W}_k[n]) \nonumber\\
 {\rm{s.t.}}  \quad & \sum_{k= 1}^{K} \mathrm{tr} (\boldsymbol{W}_k[n] ) \le P_{\max},\ \forall n\in \mathcal{N},   \label{p2a}\\
 & \mathrm{rank}(\boldsymbol{W}_k[n] ) \leq 1,\ \forall k\in \mathcal{K},\ \forall n\in \mathcal{N},\label{p2b}
\end{flalign}}\end{subequations}
where $\Tilde{r}_k[n](\boldsymbol{W}_k[n])$ is given by
\begin{equation} {\footnotesize
\begin{split}
    \Tilde{r}_k[n](\boldsymbol{W}_k[n]) = \log_2\bigg( 1+ 
        \frac{\mathrm{tr}\left( \boldsymbol{\Lambda}_k[n]\boldsymbol{W}_k[n]
        \right)}
            {\sum_{l\in\mathcal{K}\backslash\{k\}}\mathrm{tr}\left( \boldsymbol{\Lambda}_k[n]\boldsymbol{W}_l[n]
        \right)+\sigma_k^2}
    \bigg),
\end{split} }
\end{equation}
and $\boldsymbol{\Lambda}_k[n] = h_k^2[n]
        \boldsymbol{a}_k^{\mathrm{}}[n]\boldsymbol{a}_k^{\mathrm{H}}[n]$.
        
The problem (\textbf{P2}) is still non-convex regarding the non-concave nature of the objective, and the rank constraint (\ref{p2b}). Hence, we approximate the objective function by the successive convex approximation (SCA), where the $i$-th iteration of the local point of $\boldsymbol{W}_k[n]$ is $\boldsymbol{W}_k^{(i)}[n]$. The objective under $i$-th iteration can be given by
\begin{equation} {\footnotesize
\begin{split}
    &\Tilde{r}_k[n](\boldsymbol{W}_k[n]) =
    \log_2 \left(
        \sum_{l\in\mathcal{K}}\mathrm{tr}\left( \boldsymbol{\Lambda}_k[n]\boldsymbol{W}_l[n]\right) + \sigma_k^2
    \right)\\
    &\quad \ - \log_2 \left(
        \sum_{l\in\mathcal{K}\backslash\{k\}}\mathrm{tr}\left( \boldsymbol{\Lambda}_k[n]\boldsymbol{W}_l[n]
        \right)+\sigma_k^2
    \right)\\
    &\geq 
    \left(
        \sum_{l\in\mathcal{K}}\mathrm{tr}\left( \boldsymbol{\Lambda}_k[n]\boldsymbol{W}_l[n]\right) + \sigma_k^2
    \right)\\
    & - \left(
        \alpha_k^{(i)}[n] + \sum_{l\in\mathcal{K}\backslash\{k\}}\mathrm{tr}\left(
            \boldsymbol{\Delta}_k^{(i)[n]\left(\boldsymbol{W}_l[n]-\boldsymbol{W}_l^{(i)}[n] \right)}
        \right)
    \right)\\
    &\triangleq \check{r}_k[n](\boldsymbol{W}_k[n]),
\end{split}}
\end{equation}
where $\alpha_k^{(i)}[n]$ and $\boldsymbol{\Delta}_k^{(i)}[n]$ are given by
\begin{equation} {\small
\begin{split}
    \alpha_k^{(i)}[n] =
    \log_2 \left(
        \sum_{l\in\mathcal{K}\backslash\{k\}}
        \mathrm{tr} \left(
            \boldsymbol{\Lambda}_k[n]\boldsymbol{W}_l^{(i)}[n] 
        \right) + \sigma_k^2
    \right),   
\end{split} }
\end{equation}
\begin{equation}{ \small
\begin{split}
    \boldsymbol{\Delta}_k^{(i)}[n] = 
    \frac{\log_2(\mathrm{e})\boldsymbol{\Lambda}_k[n]}
    {\sum_{l\in\mathcal{K}\backslash\{k\}}
        \mathrm{tr} \left(
            \boldsymbol{\Lambda}_k[n]\boldsymbol{W}_l^{(i)}[n] 
        \right) + \sigma_k^2}.
\end{split} }
\end{equation}

Hence, the problem (\textbf{P2}) can be approximated as problem (\textbf{P2.$i$}) in the $i$-th iteration, which is given by
\begin{subequations} {\small
\begin{flalign}
 (\textbf{P2.}i):\ \max_{\{\boldsymbol{W}_k[n] \succeq 0\}} \quad & \sum_{n= 1}^{N}\sum_{k= 1}^{K}  \check{r}_k[n] (\boldsymbol{W}_k[n]) \nonumber\\
 {\rm{s.t.}}  \quad & \text{(\ref{p2a}) and (\ref{p2b})}.\nonumber
\end{flalign}}\end{subequations}
In problem (\textbf{P2.$i$}), the objective function w.r.t. $\boldsymbol{W}_k[n]$ is concave. However, the constraint (\ref{p2b}) is still non-convex. 
To address this, we relax the rank constraints to convert problem (\textbf{P2.$i$}) into a standard convex optimization problem (\textbf{RP2.$i$}). Although the relaxed solution may not ensure rank-one matrices, Gaussian randomization can approximate a rank-one solution if required. Fortunately, by leveraging similar analyses from the prior work \cite{9916163}, we find that an optimal rank-one solution $\boldsymbol{W}_k^{*}[n]$ always exists. Hence, Gaussian randomization is unnecessary for solving problem (\textbf{P2.$i$}). Due to space limitations, the detailed proof is omitted.

\subsection{Optimizing $\boldsymbol{Q}$ Given $\boldsymbol{W}$ and $\boldsymbol{X}$}
As indicated by Eq. (\ref{steering vector}), $\boldsymbol{a}(\theta_k [n])$ exhibits complexity and non-linearity w.r.t $\boldsymbol{q}[n]$, which makes the UAV trajectory optimization become very difficult. To address this issue, we utilize the $i$-th iteration of the UAV trajectory to approximate the variables in $\boldsymbol{a}(\theta_k [n])$ related to the $(i+1)$-th iteration of the UAV trajectory. Thus, $\boldsymbol{a}(\theta_k [n])$ is rewritten as
\begin{equation} {\footnotesize
    \boldsymbol{\Tilde{a}}(\theta_k [n]) = \left [
    e^{\mathrm{j}\frac{2\pi}{\lambda}x_1[n]\cos\theta_k^{i} [n]},
    \cdots,
    e^{\mathrm{j}\frac{2\pi}{\lambda}x_M[n]\cos\theta_k^{i} [n]}
    \right ]^{\mathrm{T}}, 
}\end{equation}
where $\theta_k^{i} [n] = \arccos \frac{H}{\sqrt{\Vert \boldsymbol{q}^{i}[n]-\boldsymbol{s}_k\Vert^2+H^2}}$.

Thus, the objective function of (\textbf{P1}) can be reformulated as
\begin{equation} {\footnotesize
\begin{split}
    \bar{r}_k[n](\boldsymbol{q}[n]) =
    \log_2 \left( \frac{ \Phi_{k}[n] + \Upsilon_{k}[n] }{ \boldsymbol{d}_{k}[n] } +\sigma_k^2 \right) - \log_2 \left(\frac{
    \Upsilon_{k}[n] }{ \boldsymbol{d}_{k}[n] } +\sigma_k^2  \right),
\end{split}}
\label{obb}
\end{equation}
where $\boldsymbol{d}_{k}[n] = \Vert \boldsymbol{q}[n]-\boldsymbol{s}_k\Vert^2+H^2$, $\Phi_{k}[n] = \chi \left \vert \boldsymbol{\Tilde{a}}_k^{\mathrm{H}}[n]\boldsymbol{w}_k [n] \right \vert^{2}$, and $\Upsilon_{k}[n] = \sum_{l\in\mathcal{K}\backslash\{k\}} \chi \left \vert \boldsymbol{\Tilde{a}}_k^{\mathrm{H}}[n]\boldsymbol{w}_l [n] \right \vert^{2}$. Obviously, the first term of $\bar{r}_k[n](\boldsymbol{q}[n])$ is non-concave w.r.t. $\boldsymbol{q}[n]$, but it is convex w.r.t. $\boldsymbol{d}_{k}[n]$. By applying the SCA method, the lower-bound for the first term of $\bar{r}_k[n](\boldsymbol{q}[n])$ can be expressed as
\begin{equation} {\small
\begin{split}
&\log_2 \left( \frac{ \Phi_{k}[n] + \Upsilon_{k}[n] }{ \boldsymbol{d}_{k}[n] } +\sigma_k^2 \right) \geq \log_2 \left( \frac{ \Phi_{k}[n] + \Upsilon_{k}[n] }{ \boldsymbol{d}^{i}_{k}[n] } +\sigma_k^2 \right) \\
& - \frac{ \log_2(\mathrm{e}) (\Phi_{k}[n] + \Upsilon_{k}[n]) (\boldsymbol{d}^{(i)}_{k}[n])^{-2} }{ (\Phi_{k}[n] + \Upsilon_{k}[n]) (\boldsymbol{d}^{(i)}_{k}[n])^{-1} + \sigma_k^2 } (\boldsymbol{d}_{k}[n] - \boldsymbol{d}^{(i)}_{k}[n]) \triangleq \bar{r}^{\rm{1}}_k[n].
\end{split} }
\end{equation}
Besides, the second term of (\ref{obb}) is non-convex w.r.t. $\boldsymbol{q}[n]$. By introducing a slack variable $z_{k}[n]$, this term is approximated by $\bar{r}^{\rm{2}}_k[n] \triangleq \log_2(\Upsilon_{k}[n] e^{z_{k}[n]} + \sigma_k^2)$, where $e^{z_{k}[n]}$ satisfies
\begin{equation} {\footnotesize
    e^{-z_{k}[n]} \leq \boldsymbol{d}^{(i)}_{k}[n] + 2(\boldsymbol{q}^{(i)}[n] - \boldsymbol{s}_k)^{\rm{T}}(\boldsymbol{q}[n]-\boldsymbol{q}^{(i)}[n]). \label{z_{k}[n]}}
\end{equation}
Moreover, according to the SCA method, (\ref{p1d}) can be transformed into a convex constraint as $\Vert \boldsymbol{v}[n] \Vert \leq V_{\max}$ and
\begin{equation} {\small
\begin{split}
(V_{\min} \tau)^{2} \leq & -\|\boldsymbol{q}^{(i)}[n]-\boldsymbol{q}^{(i)}[n-1]\|^2\\
    & +2(\boldsymbol{q}^{(i)}[n]-\boldsymbol{q}^{(i)}[n-1])^{\rm{T}}(\boldsymbol{q}[n]-\boldsymbol{q}[n-1]). \label{V_min}
\end{split} }
\end{equation}

Hence, the problem (\textbf{P1}) can be approximated as problem (\textbf{P3.$i$}) in the $i$-th iteration, which is given by
\begin{subequations} {\small
\begin{flalign}
 (\textbf{P3.}i):\ \max_{\boldsymbol{Q},\boldsymbol{Z}} \quad & \sum_{n= 1}^{N}\sum_{k= 1}^{K} (\bar{r}^{\rm{1}}_k[n] - \bar{r}^{\rm{2}}_k[n])  \nonumber\\
 {\rm{s.t.}}  \quad & \text{(\ref{p1e}), (\ref{z_{k}[n]}), and (\ref{V_min}),} \nonumber \\
 &\Vert \boldsymbol{v}[n] \Vert \leq V_{\max} \label{p3ia},
\end{flalign} }\end{subequations}
where $\boldsymbol{Z} = \{z_{k}[n],\forall k,\forall n \}$.
The objective of $(\textbf{P3.} i)$ and all constraints are convex w.r.t $\boldsymbol{q}[n]$, resulting the convexity of $(\textbf{P3.} i)$, Therefore, (\textbf{P3.$i$}) can be solved by standard convex optimization methods.

\subsection{Optimizing $\boldsymbol{X}$ Given $\boldsymbol{Q}$ and $\boldsymbol{W}$}
As indicated by Eq. (\ref{rate}), the achievable data rate $r_k[n]$ is a monotonically increasing function of SINR $\gamma_k[n]$. 
Therefore, the objective function of problem (\textbf{P1}) can be transformed into $\max_{\boldsymbol{X}} \sum_{n= 1}^{N}\sum_{k= 1}^{K} \gamma_k[n]$, which is a nonlinear fractional problem and can be handled by the Dinkelbach method.
By introducing a parameter $\eta_{k}[n] = \gamma_k[n]$ to denote the SINR, the objective function of (\textbf{P1}) is transformed into
\begin{equation} {\footnotesize
\begin{split}
 \Gamma_{k}[n] = &\ h_k^{2}[n] \left \vert \boldsymbol{a}_k^{\mathrm{H}}[n]
\boldsymbol{w}_k [n] \right \vert^{2} \\
&- \eta_{k}[n] \left(h_k^{2}[n] \sum_{l\in\mathcal{K}\backslash\{k\} } \vert 
    \boldsymbol{a}_k^{\mathrm{H}}[n] \boldsymbol{w}_l[n] \vert^{2} + \sigma_k^2 \right),   
    \label{seccobj1}
\end{split} }
\end{equation}
which is a non-convex function. Motivated by \cite{10382559}, we relax it by using the SCA method. For ease of exposition, we define $\vartheta_{k}[n] = \frac{2\pi}{\lambda}\cos\theta_k [n]$. Furthermore, we represent the $m$-th element of $\boldsymbol{w}_{k}[n]$ as $[{w}_{k}[n]]_{m}  =\left \vert [{w}_{k}[n]]_{m}  \right \vert e^{\mathrm{j} \angle [{w}_{k}[n]]_{m}}$, where $\left \vert [{w}_{k}[n]]_{m}  \right \vert$ denotes its amplitude and $ \angle [{w}_{k}[n]]_{m}$ denotes its phase.
Then, $\left \vert \boldsymbol{a}_k^{\mathrm{H}}[n]
\boldsymbol{w}_k [n] \right \vert^{2}$ can be further expressed as
\begin{equation} {\footnotesize
    \begin{split}
&\left \vert \boldsymbol{a}_k^{\mathrm{H}}[n]
\boldsymbol{w}_k [n] \right \vert^{2} 
= \left \vert \sum_{p=1}^{M} [{w}_{k}^{*}[n]]_{p} e^{\mathrm{j} \vartheta_{k}[n] x_{p}[n] } \right \vert^{2} \\
&\ \ \ \ = \sum_{p=1}^{M} \sum_{q=1}^{M} \left \vert [{w}_{k}[n]]_{p} [{w}_{k}[n]]_{q} \right \vert \cos \left( \Theta_{k}[n](x_{p}[n],x_{q}[n]) \right),
    \end{split} }
\end{equation}
where $ \Theta_{k}[n](x_{p}[n],x_{q}[n]) = \vartheta_{k}[n] (x_{p}[n]-x_{q}[n]) -( \angle [{w}_{k}[n]]_{p} - \angle [{w}_{k}[n]]_{q}) $.
We use Taylor expansion to deal with the $\cos\left( \Theta_{k}[n](x_{p}[n],x_{q}[n])  \right)$. Specifically, for a given $\beta \in \mathbb{R}$, the second-order Taylor
expansion of $\cos (\beta)$ is
\begin{equation} {\footnotesize
    \begin{split}
\cos (\beta) &\approx \cos (\beta_0) - \sin (\beta_0) (\beta -\beta_0) - \frac{1}{2} \cos (\beta_0) (\beta -\beta_0)^{2} \\
& \geq \cos (\beta_0) - \sin (\beta_0) (\beta -\beta_0) - \frac{1}{2} (\beta -\beta_0)^{2} \triangleq f(\beta|\beta_0).
    \end{split} }
\end{equation}
Then, for given $\boldsymbol{x}^{i}[n] = [x_1^{i}[n],\ldots,x_M^{i}[n]]^{\mathrm{T}}$ in the $i$-th iteration, by letting $\beta_0 = \Theta_{k}[n](x_{p}^{i}[n],x_{q}^{i}[n])$ and $\beta = \Theta_{k}[n](x_{p}[n],x_{q}[n])$, we can obtain
\begin{equation} {\footnotesize
\begin{split}
\left\vert \boldsymbol{a}_k^{\mathrm{H}}[n]
\boldsymbol{w}_k [n] \right \vert^{2} 
&\geq \sum_{p=1}^{M} \sum_{q=1}^{M} \vert [{w}_{k}[n]]_{p} [{w}_{k}[n]]_{q} \vert \times\\
f ( \cos ( &\Theta_{k}[n](x_{p}[n],x_{q}[n]) ) | \cos ( \Theta_{k}[n](x_{p}^{i}[n],x_{q}^{i}[n]) ) ) \\
&\triangleq \frac{1}{2} \boldsymbol{x}^{\rm{T}}[n] \boldsymbol{A}_{k}[n] \boldsymbol{x}[n] +\boldsymbol{b}_{k}^{\rm{T}}[n] \boldsymbol{x}[n]+c_{k}[n],
\end{split}}
\end{equation}
where $\boldsymbol{A}_{k}[n] \in \mathbb{R}^{M  \times M}$, $\boldsymbol{b}_{k}[n] \in \mathbb{R}^{M }$, and $c_{k}[n] \in \mathbb{R}$ are expressed as
\begin{equation} {\footnotesize
\boldsymbol{A}_{k}[n] \triangleq  -2 \vartheta_{k}^{2}[n]\left(\gamma[n] \operatorname{diag}(\Tilde{\boldsymbol{w}}_{k}[n])-\Tilde{\boldsymbol{w}}_{k}[n] \Tilde{\boldsymbol{w}}_{k}[n]^{\rm{T}}\right),}
\end{equation}
\begin{equation} {\footnotesize
    \begin{split}
& [\boldsymbol{b}_{k}[n]]_{p} \triangleq  2 \vartheta_{k}^{2}[n] \sum_{q=1}^{M}\left|[{w}_{k}[n]]_{p} [{w}_{k}[n]]_{q}\right|\left(x_{p}^{i}[n]-x_{q}^{i}[n]\right) \\
& -2 \vartheta_{k}[n] \sum_{q=1}^{M}\left|[{w}_{k}[n]]_{p} [{w}_{k}[n]]_{q}\right| \sin \left( \Theta_{k}[n](x_{p}^{i}[n],x_{q}^{i}[n]) \right),       
    \end{split}}
\end{equation}
\begin{equation} {\footnotesize
    \begin{aligned}
&c_{k}[n] \triangleq  \sum_{p=1}^{M} \sum_{q=1}^{M}\left|[{w}_{k}[n]]_{p} [{w}_{k}[n]]_{q} \right| \cos \left( \Theta_{k}[n](x_{p}^{i}[n],x_{q}^{i}[n]) \right) \\
& +\vartheta_{k}[n] \sum_{p=1}^{M} \sum_{q=1}^{M}\left|[{w}_{k}[n]]_{p} [{w}_{k}[n]]_{q}\right| \sin \left( 
 \Theta_{k}[n](x_{p}^{i}[n],x_{q}^{i}[n]) \right) \times \\
 & \left(x_{p}^{i}[n]-x_{q}^{i}[n]\right) 
 -\frac{1}{2} \vartheta_{k}^{2}[n] \sum_{p=1}^{M} \sum_{q=1}^{M}\left|[{w}_{k}[n]]_{p} [{w}_{k}[n]]_{q}\right| \times \\
& \left(x_{p}^{i}[n]-x_{q}^{i}[n]\right)^{2},  
    \end{aligned}}
\end{equation}
with $\Tilde{\boldsymbol{w}}_{k}[n] \triangleq [ [{w}_{k}[n]]_{1}, [{w}_{k}[n]]_{2}, \ldots, [{w}_{k}[n]]_{M} ]^{\rm{T}} $ and $\gamma[n] \triangleq \sum_{p=1}^{M} | [{w}_{k}[n]]_{p}| $.
By introducing a slack variable $\delta_{k}[n]$, we can obtain a standard quadratic constraint, i.e.,
\begin{equation} {\footnotesize
    \frac{1}{2} \boldsymbol{x}^{\rm{T}}[n] \boldsymbol{A}_{k}[n] \boldsymbol{x}[n] +\boldsymbol{b}_{k}^{\rm{T}}[n] \boldsymbol{x}[n]+c_{k}[n] \geq \delta_{k}[n]. \label{R1_low}}
\end{equation}
Moreover, given that $\left \vert \boldsymbol{a}_k^{\mathrm{H}}[n]
\boldsymbol{w}_l [n] \right \vert^{2}$ has a structure similar to $\left \vert \boldsymbol{a}_k^{\mathrm{H}}[n]
\boldsymbol{w}_k [n] \right \vert^{2}$, we can transform it by modifying the procedure utilized to construct the relaxed quadratic constraint (\ref{R1_low}). Specifically, we have
\begin{equation} {\footnotesize
    \begin{split}
\cos (\beta) &\approx \cos (\beta_0) - \sin (\beta_0) (\beta -\beta_0) - \frac{1}{2} \cos (\beta_0) (\beta -\beta_0)^{2} \\
& \leq \cos (\beta_0) - \sin (\beta_0) (\beta -\beta_0) + \frac{1}{2} (\beta -\beta_0)^{2} \triangleq \Tilde{f}(\beta|\beta_0),
    \end{split}}
\end{equation}
\begin{equation} {\footnotesize
\begin{split}
\left\vert \boldsymbol{a}_k^{\mathrm{H}}[n]
\boldsymbol{w}_l [n] \right\vert^{2}
&\leq \sum_{p=1}^{M} \sum_{q=1}^{M} \vert [{w}_{l}[n]]_{p} [{w}_{l}[n]]_{q} \vert \times\\
\Tilde{f} ( \cos ( \Tilde{\Theta}_{k,l}&[n](x_{p}[n],x_{q}[n]) ) \vert \cos ( \Tilde{\Theta}_{k,l}[n](x_{p}^{i}[n],x_{q}^{i}[n]) ) ) \\
& \triangleq \frac{1}{2} \boldsymbol{x}^{\rm{T}}[n] \Tilde{\boldsymbol{A}}_{k,l}[n] \boldsymbol{x}[n] +\Tilde{\boldsymbol{b}}_{k,l}^{\rm{T}}[n] \boldsymbol{x}[n] + \Tilde{c}_{k,l}[n].
\end{split}}
\end{equation}
where $\Tilde{\Theta}_{k,l}[n](x_{p}[n],x_{q}[n]) = \vartheta_{k}[n] (x_{p}[n]-x_{q}[n]) -( \angle [{w}_{l}[n]]_{p} - \angle [{w}_{l}[n]]_{q})$, $\Tilde{\boldsymbol{A}}_{k.l}[n] \in \mathbb{R}^{M  \times M}$, $\Tilde{\boldsymbol{b}}_{k,l}[n] \in \mathbb{R}^{M }$, and $\Tilde{c}_{k,l}[n] \in \mathbb{R}$ are expressed as
\begin{equation} {\footnotesize
\Tilde{\boldsymbol{A}}_{k,l}[n] \triangleq  2 \vartheta_{k}^{2}[n]\left(\Tilde{\gamma}[n] \operatorname{diag}(\Tilde{\boldsymbol{w}}_{l}[n])-\Tilde{\boldsymbol{w}}_{l}[n] \Tilde{\boldsymbol{w}}_{l}[n]^{\rm{T}}\right),}
\end{equation}
\begin{equation} {\footnotesize
    \begin{split}
& [\Tilde{\boldsymbol{b}}_{k,l}[n]]_{p} \triangleq  -2 \vartheta_{k}^{2}[n] \sum_{q=1}^{M}\left|[{w}_{l}[n]]_{p} [{w}_{l}[n]]_{q}\right|\left(x_{p}^{i}[n]-x_{q}^{i}[n]\right) \\
& -2 \vartheta_{k}[n] \sum_{q=1}^{M}\left|[{w}_{l}[n]]_{p} [{w}_{l}[n]]_{q}\right| \sin \left( \Tilde{\Theta}_{k,l}[n](x_{p}^{i}[n],x_{q}^{i}[n]) \right),       
    \end{split}}
\end{equation}
\begin{equation} {\footnotesize
    \begin{split}
&\Tilde{c}_{k,l}[n] \triangleq  \sum_{p=1}^{M} \sum_{q=1}^{M}\left|[{w}_{l}[n]]_{p} [{w}_{l}[n]]_{q} \right| \cos \left( \Tilde{\Theta}_{k,l}[n](x_{p}^{i}[n],x_{q}^{i}[n]) \right) \\
& +\vartheta_{k}[n] \sum_{p=1}^{M} \sum_{q=1}^{M}\left|[{w}_{l}[n]]_{p} [{w}_{l}[n]]_{q}\right| \sin \left( 
 \Tilde{\Theta}_{k,l}[n](x_{p}^{i}[n],x_{q}^{i}[n]) \right) \times \\
 & \left(x_{p}^{i}[n]-x_{q}^{i}[n]\right) 
 +\frac{1}{2} \vartheta_{k}^{2}[n] \sum_{p=1}^{M} \sum_{q=1}^{M}\left|[{w}_{l}[n]]_{p} [{w}_{l}[n]]_{q}\right| \times \\
& \left(x_{p}^{i}[n]-x_{q}^{i}[n]\right)^{2},  
    \end{split}}
\end{equation}
with $\Tilde{\gamma}[n] =  \sum_{p=1}^{M} | [{w}_{l}[n]]_{p}| $.
By introducing a slack variable $\zeta_{k,l}[n]$, we can obtain a standard quadratic constraint, i.e.,
\begin{equation} {\footnotesize
 \frac{1}{2} \boldsymbol{x}^{\rm{T}}[n] \Tilde{\boldsymbol{A}}_{k,l}[n] \boldsymbol{x}[n] +\Tilde{\boldsymbol{b}}_{k,l}^{\rm{T}}[n] \boldsymbol{x}[n] + \Tilde{c}_{k,l}[n]\leq \zeta_{k,l}[n]. \label{R2_up}}
\end{equation}
which can be solved efficiently by convex optimization solvers. Hence, the objective function (\ref{seccobj1}) can be converted to
\begin{equation} {\footnotesize
\begin{split}
  \Tilde{\Gamma}_{k}[n] = \left( h_k^{2}[n] \delta_{k}[n] - \eta_{k}[n] \left( h_k^{2}[n] \sum_{l\in\mathcal{K}\backslash\{k\} } \zeta_{k,l}[n]  + \sigma_k^2 
 \right)\right).
\end{split} }
\end{equation}
Then, the problem (\textbf{P1}) can be approximated as problem (\textbf{P4.$i$}) in the $i$-th iteration, which is given by
\begin{subequations} \small
\begin{flalign}
 (\textbf{P4.}i):\ \max_{\boldsymbol{Q},\boldsymbol{\delta},\boldsymbol{ \zeta}} \quad  \sum_{n= 1}^{N}\sum_{k= 1}^{K} \Tilde{\Gamma}_{k}[n]  \quad \
 {\rm{s.t.}}  \quad  \text{(\ref{p1b}), (\ref{p1c}), (\ref{R1_low}), and (\ref{R2_up}),} \nonumber
\end{flalign}
\end{subequations}
where $\boldsymbol{\delta}=\{\delta_{k}[n]\}$ and $\boldsymbol{\zeta}=\{  \zeta_{k,l}[n]\}$.
Problem (\textbf{P4.$i$}) is reformulated as a standard convex problem w.r.t to $\boldsymbol{X}$, and can be solved by CVX toolbox.

\subsection{Alternating Optimization}
The AO-based algorithm for solving (\textbf{P1}) is depicted in Algorithm \ref{Alg}\footnote{
{The solution of $\boldsymbol{W}$ and $\boldsymbol{X}$ are independent across different time slots. To reduce the complexity of the algorithm, we decompose problem (\textbf{P2.}$i$) and problem (\textbf{P4.}$i$) into $N$ subproblems. Each subproblem is then optimized independently within its respective time slot.
}}. Given that the total achievable data rate is non-decreasing over iterations and has an upper bound, Algorithm \ref{Alg} is ensured to converge. In addition, the aforementioned three sub-problems are solved by interior point method through CVX.
The complexity of the three sub-problems is $\mathcal{O} ((NKM^{2})^{3.5} \log(\epsilon_{\epsilon^*}^{-1}))$, $\mathcal{O} ((2N)^{3.5} \log(\epsilon_{\epsilon^*}^{-1}))$, and $\mathcal{O} ((MN)^{3.5} \log(\epsilon_{\epsilon^*}^{-1}))$, respectively. As a result, the overall complexity of solving (\textbf{P1}) is $\mathcal{O} ( i_{\max} ( ( (NKM^{2})^{3.5} +(2N)^{3.5} +(MN)^{3.5}) \log(\epsilon_{\epsilon^*}^{-1})  )$.

\begin{algorithm} \scriptsize
	\caption{AO for Solving (\textbf{P1})}
	\label{Alg}
	\begin{algorithmic}
		
		\REQUIRE {An initial feasible point $\boldsymbol{W}^{(0)}$, $\boldsymbol{Q}^{(0)}$, and $\boldsymbol{X}^{(0)}$;}\\
		\textbf{Initialize:} the iteration number $i=1$;
		the precision threshold $\epsilon^*$;
            the maximum number of iterations $i_{\max}$;\\
		
		\REPEAT
		\STATE Given $\{\boldsymbol{Q}^{(i-1)}, \boldsymbol{X}^{(i-1)}\}$ to solve the problem \textbf{P2}.$i$, get the solution $\boldsymbol{W}^{(i)}$;\\
		\STATE Given $\{\boldsymbol{W}^{(i)}, \boldsymbol{X}^{(i-1)}\}$ to solve the problem \textbf{P3}.$i$, get the solution $\boldsymbol{Q}^{(i)}$;\\
            \STATE Given $\{\boldsymbol{W}^{(i)}, \boldsymbol{Q}^{(i)}\}$ to solve the problem \textbf{P4}.$i$, get the solution $\boldsymbol{X}^{(i)}$;\\
		\STATE Set $i=i+1$; \\
		\UNTIL {the gain of objective $\leq \epsilon^*$ or $i = i_{\max}$};\\
		\ENSURE {$\boldsymbol{W}^{*}[n]$, $\boldsymbol{Q}^{*}[n]$, and $\boldsymbol{X}^{*}[n]$.}\\
	\end{algorithmic}
\end{algorithm}

\section{Numerical Results}
In this section, we provide numerical results to validate the performance of our proposed scheme. In the simulation, we consider an area of $500 \mathrm{m} \times 500 \mathrm{m} $ with $K = 3$ users. Similar to \cite{10416363, 9916163}, we set the minimum distance between adjacent MAs as $d_{\min} = \lambda /2$, the range of positions of all antennas as $L = 8 \lambda$, the maximum transmit power as $P_{\max} = 3 \rm{W}$, the noise power at each receiver as $\sigma_k^2 = -110 \rm{dBm}$, the constant value $\chi = -60 \rm{dB}$, the range of UAV’s horizontal flight speed as $[V_{\min},V_{\max}] = [1,20] \rm{m/s}$, the flight altitude as $H = 100 \rm{m}$, and the time period $T = 40 \rm{s}$.

\begin{figure}[!h]
	\begin{minipage}{0.495\linewidth}
		\centering
		\includegraphics[width=1\linewidth]{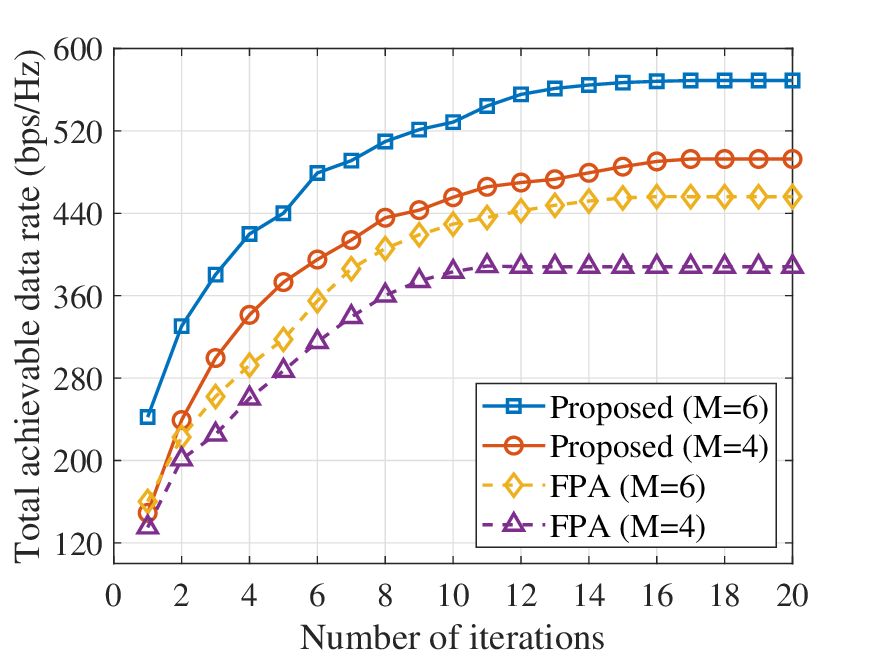}
		\captionsetup{font=small}
            \caption{\centering{Convergence.
            }}
		\label{convergence}
	\end{minipage}
	\begin{minipage}{0.495\linewidth}
		\centering
		\includegraphics[width=1\linewidth]{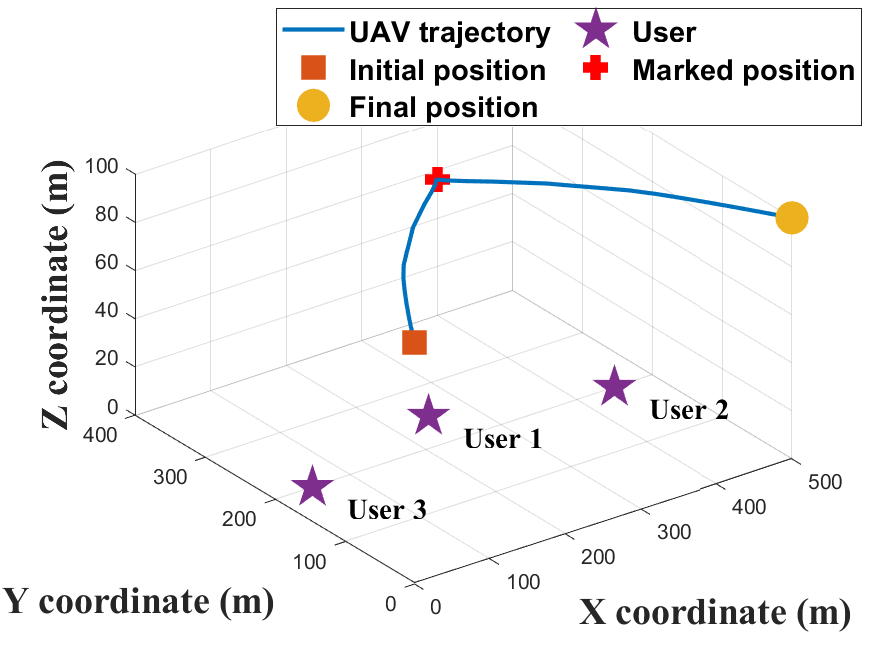}
            \captionsetup{font=small}
		\caption{\centering{3D UAV trajectory of the proposed algorithm.
            }}
		\label{UAV_trajectory}
	\end{minipage}
\end{figure}

\begin{figure}[!h]
	\begin{minipage}{0.495\linewidth}
		\centering
		\includegraphics[width=1\linewidth]{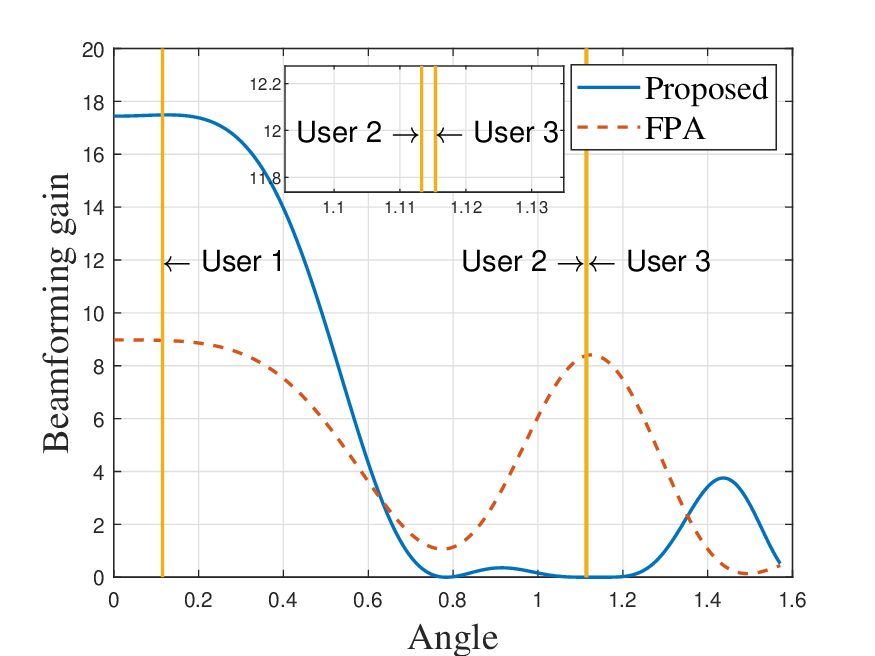}
		\captionsetup{font=small}
            \caption{\centering{Comparison of beamforming gains with MA array and FPA array.}}
		\label{beamforming_gain1}
	\end{minipage}
	\begin{minipage}{0.495\linewidth}
		\centering
		\includegraphics[width=1\linewidth]{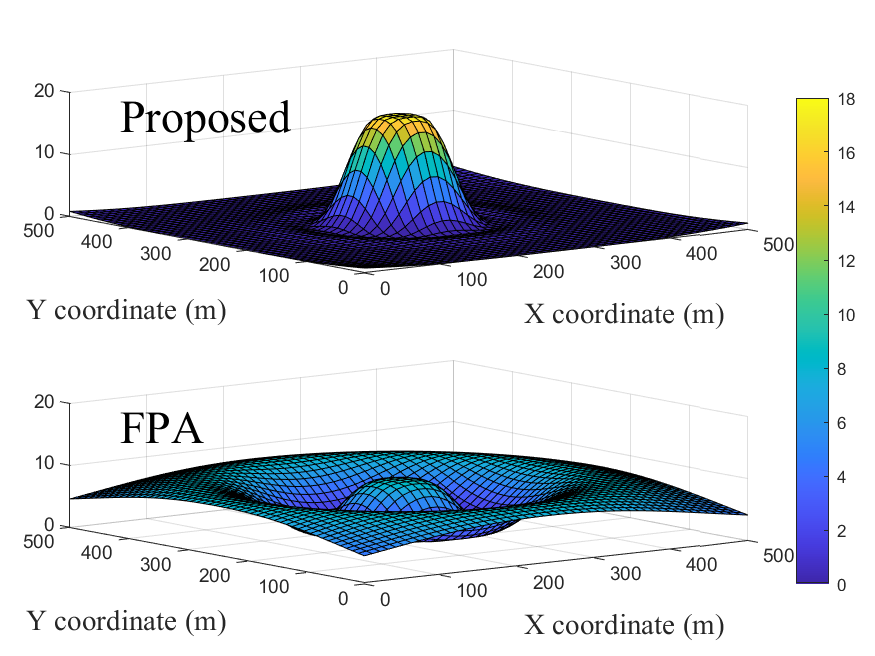}
            \captionsetup{font=small}
		\caption{\centering{Comparison of beamforming gains with MA array and FPA array in 2D space.
            }}
		\label{beamforming_gain2}
	\end{minipage}
\end{figure}

The convergence of the proposed MA array-based scheme and the 
\textit{fixed-position antenna array} (FPA) scheme (only the beamforming and the trajectory are optimized) is illustrated in Fig.~\ref{convergence}.
The proposed AO-based algorithm is observed to be efficient where all schemes can converge within $16$ iterations. The proposed scheme with $M=6$ achieves the highest total achievable data rate due to the fully optimized beamforming gain with more antennas. An example of optimized UAV trajectory is demonstrated in Fig.~\ref{UAV_trajectory}. We can observe that the UAV flies to the suitable positions between users to achieve better performance.
{Note that we analyse the beamforming gain in Fig.~\ref{beamforming_gain1} and Fig.~\ref{beamforming_gain2} at the marked UAV position (the red $\{+\}$ marked in Fig.~\ref{UAV_trajectory}) toward the three users.}
\begin{figure}[!h]
	\begin{minipage}{0.495\linewidth}
		\centering
		\includegraphics[width=1\linewidth]{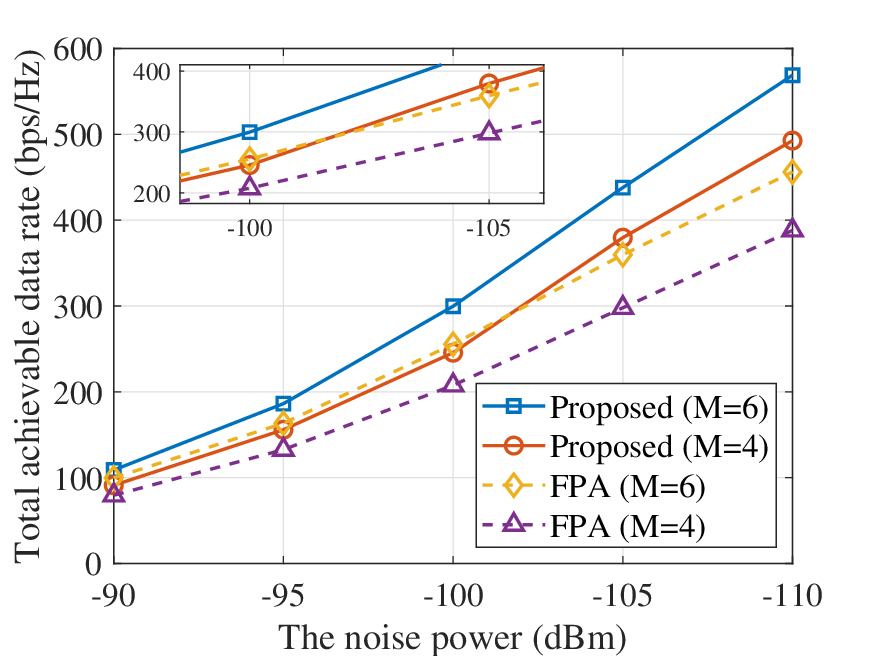}
		\captionsetup{font=small}
            \caption{\centering{The achievable data rate versus noise power.}}
		\label{noise_power_vs_rate}
	\end{minipage}
	\begin{minipage}{0.495\linewidth}
		\centering
		\includegraphics[width=1\linewidth]{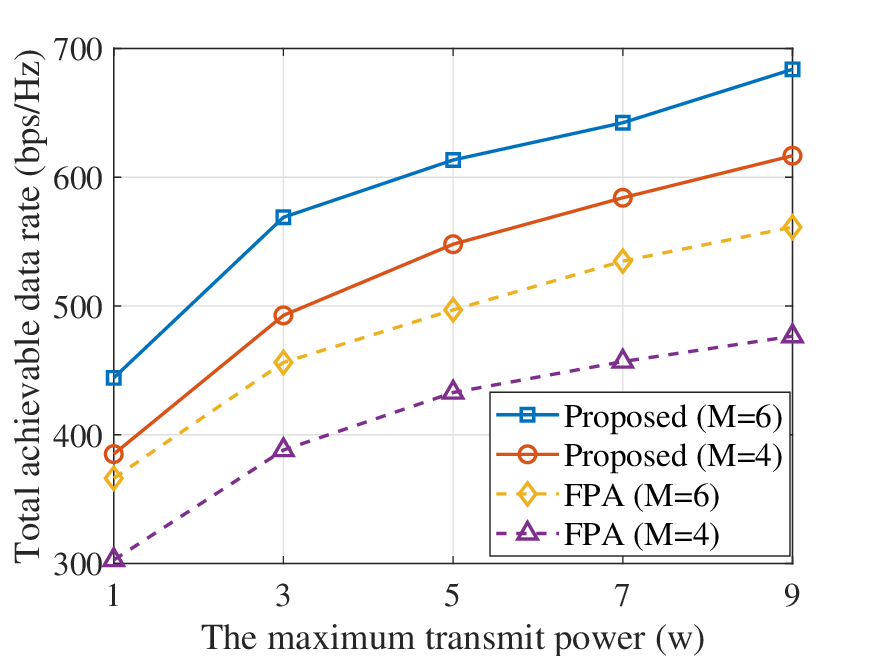}
            \captionsetup{font=small}
		\caption{\centering{The achievable data rate versus transmit power.
            }}
		\label{transmit_power_vs_rate}
	\end{minipage}
\end{figure}

In Fig.~\ref{noise_power_vs_rate}, the total achievable data rate versus the noise power is presented.
As we can observe, the proposed scheme with fewer antennas (the red curve where $M=4$) even achieves higher performance compared to the FPA scheme with a larger number of antennas (the yellow curve where $M=6$) under low noise power ($\leq -105\mathrm{dBm}$), which verifies the performance gain benefited from the optimization of antenna positions via MA array.
As shown in Fig.~\ref{transmit_power_vs_rate}, the total achievable data rate versus the maximum transmit power is illustrated.
From Fig.~\ref{transmit_power_vs_rate}, the performance gap between the proposed scheme and the FPA scheme with same number of antennas increases with the maximum transmit power, which demonstrates the effectiveness of beamforming gain from the optimization of antenna positions via MA array.

Fig.~\ref{beamforming_gain1} and Fig.~\ref{beamforming_gain2} show the beamforming gain of the proposed scheme and the FPA scheme at the angle $\theta$, where $\theta \in [0, \pi/2]$. The UAV is flying at the marked position denoted in Fig.~\ref{UAV_trajectory}.
{We can observe that the proposed MA scheme allocates more beamforming gain towards User $1$.
The reason is that, the proposed MA scheme is capable of adjusting the antenna position dynamically according to the positions of the UAV and the users to enable the flexible beamforming. Therefore, higher performance is guaranteed.}

\section{Conclusion}
This letter explores the total achievable data rate maximization problem in a novel UAV-enabled MU-MISO systems enhanced by the MA array. The transmit beamforming, the UAV trajectory, and the antenna positions of MA array are jointly optimized by an AO-based algorithm.
Numerical results verify the effective performance gain obtained by the MA array, compared with fixed-position antenna array scenario.

\bibliographystyle{IEEEtran}
\bibliography{myref}
\end{document}